\documentstyle[sprocl]{article}

\input amssym.def
\input amssym

\def\Journal#1#2#3#4{{#1} {\bf #2}, #3 (#4)}

\def\CQG{\em Class. Quantum Grav.}
\def\GRG{\em Gen. Rel. Grav.}

\newcommand{\etal}{{\em et al}}
                                          
\newcommand{\ie}{i.e.,}

\newcommand{\dofs}{degrees of freedom}

\newcommand{\wrt}{with respect to}
\newcommand{\gr}{general relativity}
\newcommand{\hog}{higher-order gravity}

\newcommand{\fn}{\frak{n}}
\newcommand{\Fh}{\frak{H}}
\newcommand{\Fl}{\frak{L}}

\newcommand{\Flb}{\underline{\Fl}}

\newcommand{\ch}{{\cal H}}
\newcommand{\cl}{{\cal L}}
\newcommand{\cm}{{\cal M}}
\newcommand{\cp}{{\cal P}}
\newcommand{\cq}{{\cal Q}}

\newcommand{\rmc}{{\rm{c}}}
\newcommand{\rmd}{{\rm{d}}}
\newcommand{\rme}{{\rm{e}}}

\newcommand{\Lie}{\cl_{\vec{n}}}
\newcommand{\emb}{{^{(3)}} \!}

\begin{document}

\title{HAMILTONIAN FORMULATION AND EXACT SOLUTIONS OF BIANCHI TYPE--I MODEL
       IN CONFORMAL GRAVITY}

\author{L. QUERELLA}

\address{Institut d'Astrophysique et de G\'{e}ophysique, Universit\'{e} de 
         Li\`{e}ge \\
         Avenue de Cointe 5, B-4000 Li\`{e}ge, Belgium \\
         E-mail: L.Querella@ulg.ac.be}

\maketitle
\abstracts{We develop a Hamiltonian formulation of Bianchi type--I 
cosmological model in conformal gravity, \ie{} the theory described by the 
Lagrangian $\Fl = C_{abcd} C^{abcd}$, which involves the quadratic curvature 
invariant constructed from the Weyl tensor, in a four-dimensional spacetime. 
We derive the explicit forms of the super-Hamiltonian and the constraint 
expressing the conformal invariance of the theory, and we write down the 
system of canonical equations. To seek out exact solutions to this system we 
add extra constraints on the canonical variables and we go through a 
{\em global involution algorithm} that possibly leads to the closure of the 
constraint algebra. This enables us to extract all possible particular 
solutions that may be written in closed analytical form. On the other hand,
probing the local analytical structure we show that the system does not 
possess the Painlev\'{e} property (presence of movable logarithms) and that it 
is therefore not integrable. We stress that there is a very fruitful interplay 
of {\em local integrability-related methods} such as the Painlev\'{e} test and 
global techniques such as the involution algorithm. Strictly speaking, we 
demonstrate that the global involution algorithm has proven to be exhaustive 
in the search for exact solutions. The conformal relationship of the 
solutions, or absence thereof, with Einstein spaces is highlighted.}

\section{Introduction}
\label{sec:intro}

Unlike what happens in \gr, where unwanted second-order derivative terms of 
the metric can be discarded from the gravitational action through pure 
divergences, there is no chance whatsoever to weed out such terms in the 
context of \hog{} theories. It is fortunate however that a consistent method 
of building up a Hamiltonian formulation of those theories does actually 
exist; it is a generalization of the classical {\em Ostrogradsky formalism}.~%
\cite{gitma} Basically it consists in introducing auxiliary \dofs{} that 
encompass each of the successive derivative terms higher than first order. 
Without resorting explicitly to this method Boulware worked out a Hamiltonian 
formulation of quadratic gravity.~\cite{boulw} Contradistinctively, Buchbinder 
and Lyakhovich developed a canonical formalism for the most general quadratic 
gravitational Lagrangian in four dimensions, by employing the aforementioned 
generalized Ostrogradsky method.~\cite{buchb} In a previous work we have 
already applied Boulware's canonical formalism to Bianchi cosmologies, for the 
pure $R^2$ variant of the general quadratic theory.~\cite{demar} In this
contribution we consider the conformally invariant case, which is based on the 
quadratic Lagrangian density $\Fl = \sqrt{-g} \, C_{abcd} C^{abcd}$, where 
$C^a_{\ bcd}$ is the Weyl tensor.%
\footnote{We adopt Wald's conventions of sign and definitions of curvature 
          tensors (but the extrinsic curvature).~\cite{wald}} 
Making use of a slightly different generalized Ostrograsky construction (as
compared to Buchbinder and Lyakhovich's formalism) we derive the canonical 
form of the conformally invariant action, which was first obtained by 
Boulware. We further particularize the Hamiltonian formalism to the Bianchi 
type--I cosmological model and write down the canonical equations. This is our 
starting point for seeking out exact particular solutions and for asking 
whether the conformal Bianchi type--I cosmological model is integrable or not.
In that respect, we sum up very recent results which can be found elsewhere.~%
\cite{jdlqcs}

\section{Conformal gravity and Bach equations}
\label{sec:bach}

\noindent
Consider a four-dimensional spacetime $(\cm ,g)$ and the quadratic 
gravitational action
\begin{equation} \label{ConfAct1}
   S = - \frac14 \int_{\cm} \rmd^4 x \, \sqrt{-g} \, C_{abcd} C^{abcd}, 
\end{equation}
where $C_{abcd}$ is the Weyl tensor. Its variation \wrt{} the metric yields 
the conformally invariant fourth-order equations
\begin{equation} \label{BachEq1}
   B_{ab} := 2 \nabla^m \nabla^n C_{mabn} + C_{mabn} R^{mn} = 0,
\end{equation}
which have been put up by Bach~\cite{bach} who adopted Weyl's paradigm of a 
conformally invariant gravitational theory,~\cite{weyl} though without 
considering the additional Weyl 1-form. In Eq.~(\ref{BachEq1}), $B_{ab}$ is 
called the Bach tensor; it is symmetric, trace-free and conformally invariant 
of weight $-1$. Although Eq.~(\ref{BachEq1}) is a very compact formula it is
not appropriate for calculational purpose with computer algebra --- the 
contracted double covariant derivative of the Weyl tensor is rather heavy to 
compute even for simple metrics. Recently Tsantilis \etal\/ have provided an 
algorithm for the {\sc MathTensor} package that gives the Bach equations in a 
much more tractable form, especially with regard to cosmological applications,  
and which is based on the decomposition of the Riemann tensor in its 
irreducible pieces.~\cite{tsant} In accordance with their results we write 
down the following equivalent expression of $B_{ab}$, as given by 
Eq.~(\ref{BachEq1}), 
\begin{equation} \label{BachEq2}
   B_{ab} = - \Box \left( R_{ab} - \frac{R}6 g_{ab} \right) 
            + \frac13 \nabla_a \nabla_b R
            + \left( C_{mabn} + R_{mabn} + R_{mb} g_{an} \right) R^{mn}.
\end{equation}

The simplest cosmological model exhibiting non trivial physical \dofs{} in the 
conformally invariant gravitational theory based on the action given in Eq.~%
(\ref{ConfAct1}) is the spatially homogeneous anisotropic Bianchi type--I 
model. (The isotropic {\sc flrw} cosmological models are conformally flat.) 
Writing the metric in such a way that the conformal invariance becomes 
manifest already from the outset,
\begin{equation} \label{Metric1}    
   \rmd s^2 = \rme^{2 \mu} \left[ 
                              - \rmd t^2 
                              + \rme^{2 \left( 
                                           \beta_+ + \sqrt{3} \beta_- 
                                        \right)} \rmd x^2
                              + \rme^{2 \left( 
                                           \beta_+ - \sqrt{3} \beta_- 
                                        \right)} \rmd y^2
                              + \rme^{- 4 \beta_+} \rmd z^2
                           \right], 
\end{equation}
the corresponding Bach equations, as given in Eq.~(\ref{BachEq2}), can be 
derived with the help of symbolic computational packages such as the 
{\sc excalc} package in {\sc reduce}.~\cite{jdlqcs}

\section{Hamiltonian Bianchi type--I cosmology}
\label{sec:ham}

\subsection{Hamiltonian formalism and canonical equations}

The conformally invariant action in Eq. (\ref{ConfAct1}) can be cast into 
Hamiltonian form by means of a generalized Ostrogradsky construction.

Assume first that spacetime is foliated into a family of Cauchy hypersurfaces 
$\Sigma_t$ of unit normal $n^a$. The induced metric $h_{ab}$ onto these 
hypersurfaces is defined by the formula $h_{ab}=g_{ab}+n_a n_b$. The way the 
hypersurfaces are embedded into spacetime is provided by the extrinsic 
curvature tensor $K_{ab}:= - \frac12 \Lie h_{ab}$, where $\Lie$ denotes the 
Lie derivative along the normal $n^a$ (we adopt $\Lie$ as a generalized notion 
of time differentiation). The standard {\sc adm} variables are introduced: the
{\em lapse function} $N$ and {\em shift vector} $N^i$. The $3+1$--splitting of 
spacetime enables us to express the Lagrangian density 
$\Fl= -\frac14 \sqrt{-g} \, C_{abcd} C^{abcd}$ only in terms of the quantities 
defined onto $\Sigma_t$. After some algebra we obtain the following equations,
\wrt{} the {\sc adm} basis:%
\footnote{Latin indices $i,j,k,\dots$ refer to spacelike components in 
          $\Sigma_t$; the {\sc adm} basis is defined by 
          $\vec{e}_{\fn}:=\vec{n}=(\partial_t - N^i \partial_i)/N$ and 
          $\vec{e}_i = \partial_i$; a stroke `$\mid$' denotes the covariant 
          derivative onto $\Sigma_t$, and the symbol $\fn$ indicates a 
          component along $\vec{e}_{\fn}$.} 
\begin{eqnarray} 
   C^{\fn}_{\ ijk} &=& 
      \left[ 
         \delta_i^r \delta_j^s \delta_k^t - 
         \frac12 h^{rt} \left( h_{ik} \delta_j^s - h_{ij} \delta_k^s \right)
      \right] 
      \left( K_{rs\mid t} - K_{rt\mid s} \right), \label{WeylADM1a} \\
   C_{\fn i \fn j} &=&
      \frac12 \left( \delta_i^k \delta_j^l - \frac13 h_{ij} h^{kl} \right)
              \left( 
                 \Lie K_{kl} + \frac{N_{\mid kl}}{N} + K K_{kl} + \emb{R}_{kl}
              \right), \label{WeylADM1b}
\end{eqnarray}
Owing to the fact that the Weyl tensor identically vanishes 
in three dimensions, the Lagrangian density in Eq.~(\ref{ConfAct1}) reduces to
\begin{equation} \label{WeylADM2}
   \Fl = - N \sqrt{h} \, 
         \left( 
            2 C_{\fn i \fn j} C^{\fn i \fn j} + 
              C_{\fn ijk} C^{\fn ijk} 
         \right),
\end{equation}
where $C_{\fn ijk}$ and $C_{\fn i \fn j}$ are given by Eq.~(\ref{WeylADM1a}) 
and Eq.~(\ref{WeylADM1b}), respectively. 

Now consider that the induced metric $h_{ij}$ and the extrinsic curvature
$K_{ij}$ are {\em independent} variables --- \ie{} $K_{ij}$ are introduced as 
auxiliary Ostrogradsky variables. In order to recover the definition 
$K_{ij}= - \frac12 \Lie h_{ij}$ we must trade the original Lagrangian density
for a constrained Lagrangian density
\begin{equation} \label{ExtLag}
   \Flb = N^{-1} \Fl + \lambda^{ij} \left( \Lie h_{ij} + 2 K_{ij} \right),
\end{equation}
with Lagrange multipliers $\lambda^{ij}$ as additional variables. Thus we must 
resort to Dirac's formalism for constrained systems.~\cite{henne} The 
generalized Ostrogradsky construction enables us to cast the action in Eq.~%
(\ref{CanAct1}) into canonical form,
\begin{equation} \label{CanAct1}
   S = \int_{\cm} \rmd^4 x \, N 
          \left[ 
             p^{ij} \, \Lie h_{ij} + \cq^{ij} \, \Lie K_{ij} -
             \Fh_{\rmc} \left( h, K, p, \cq \right)
          \right],  
\end{equation}
with the conjugate momenta $p^{ij}= \lambda^{ij}$ and 
$\cq^{ij}= - 2 \sqrt{h} \, C^{\fn i \fn j}$, and where the canonical 
Hamiltonian density is given by
\begin{equation} \label{CanHam}
   \Fh_{\rmc} = - 2 p^{ij} K_{ij} + \sqrt{h} \, C_{\fn ijk} C^{\fn ijk} 
                - \frac{\cq^{ij} \cq_{ij}}{2 \sqrt{h}}  
                - \cq^{ij}_{\ \ \mid ij} 
                - \cq^{ij} \emb{R}_{ij} 
                - K K_{ij} \cq^{ij}.
\end{equation}
Dirac's constraint analysis yields, besides the usual super-Hamiltonian and
super-momentum constraints, one first-class constraint that is the generator 
of conformal transformations. It reads explicitly
\begin{equation} \label{SecCon}
   \chi = 2 p + K_{kl} \cq^{kl} \approx 0.
\end{equation}
Moreover we can get rid of the spurious canonical variables $\cq$ and $K$; 
only the traceless part of the corresponding tensors remain as relevant 
canonical variables. This is consistent with the fact that conformal gravity
exhibits six \dofs.

Performing a canonical transformation that we have defined in a previous work~%
\cite{demar} in order to disentangle terms stemming respectively from the pure 
$R^2$ and conformal variants of the general quadratic theory, we get 
\begin{equation} \label{CanAct2}
   S = \int_{\cm} \! \rmd^4 x 
          \left[ 
             \Pi_{\mu} \dot{\mu} + \Pi_+ \dot{\beta}_+ + \Pi_- \dot{\beta}_- + 
             \cq_+ \dot{\cp}_+ + \cq_- \dot{\cp}_- - 
             N \ch_{\rmc} - \lambda_{\rmc} \varphi_{\rmc}
          \right],     
\end{equation}   
where the first-class constraints $\ch_{\rmc} \approx 0$ (super-Hamiltonian)
and $\varphi_{\rmc} \approx 0$ are given respectively by the following 
expressions
\begin{eqnarray}
   \ch_{\rmc} &=& - \frac1{\sqrt{6}} 
                      \left[ 
                         \Pi_+ \cp_+ + \Pi_- \cp_- + 
                         2 \cq_+ \left( \cp^2_+ - \cp^2_- \right) -
                         4 \cp_+ \cp_- \cq_-                                              
                      \right] \nonumber \\ 
              & & \quad - \frac12 \rme^{- 3 \mu} 
               \left( \cq^2_+ + \cq^2_- \right), \label{superHam} \\ 
   \varphi_{\rmc} &=& \Pi_{\mu} - \cp_+ \cq_+ - \cp_- \cq_-. \label{confcons}
\end{eqnarray}
A suitable gauge-fixing condition that eliminates variables $\mu$ and
$\Pi_{\mu}$, and the choice $N=\rme^{\mu}$ yield the final form of the 
canonical action
\begin{equation} \label{CanAct3}
   S = \int_{\cm} \! \rmd^4 x \, 
          \left[ 
             \Pi_+ \dot{\beta}_+ + \Pi_- \dot{\beta}_- + 
             \cq_+ \dot{\cp}_+ + \cq_- \dot{\cp}_- - \ch_{\rmc} 
          \right],     
\end{equation}   
where the super-Hamiltonian is now given by the following expression
\begin{equation} \label{superHam2}
   \ch_{\rmc} = \frac1{\sqrt{6}} 
                      \left[ 4 \cp_+ \cp_- \cq_- -
                             \Pi_+ \cp_+ - \Pi_- \cp_- - 
                         2 \cq_+ \left( \cp^2_+ - \cp^2_- \right)      
                      \right]
                 - \frac12 \left( \cq^2_+ + \cq^2_- \right).  
\end{equation}
Varying the action as given by Eq.~(\ref{CanAct3}) with respect to the 
remaining canonical variables and their conjugate momenta, we obtain the 
canonical equations for the Bianchi type--I model in conformal gravity
\footnote{To enable a straightforward comparison with Boulware's formalism we
also have performed a canonical transformation that interchanges the 
coordinates and momenta $\cp_{\pm}$ and $\cq_{\pm}$.} 
\begin{eqnarray}
   \dot{\beta}_{\pm} &=& - \frac1{\sqrt{6}} \cp_{\pm}, \label{CanEq1} \\ 
   \dot{\Pi}_{\pm}   &=& 0, \label{CanEq2} \\
   \dot{\cq}_+       &=& - \frac1{\sqrt{6}} 
                           \left( \Pi_+ + 4 \cp_- \cq_- - 4 \cp_+ \cq_+ \right), 
                           \label{CanEq3} \\
   \dot{\cq}_-       &=& - \frac1{\sqrt{6}} 
                           \left( \Pi_- + 4 \cp_- \cq_+ + 4 \cp_+ \cq_- \right), 
                           \label{CanEq4} \\
   \dot{\cp}_+       &=&   \frac2{\sqrt{6}} \left( \cp^2_- - \cp^2_+ \right) + 
                           \cq_+, \label{CanEq5} \\         
   \dot{\cp}_-       &=&   \frac4{\sqrt{6}} \cp_+ \cp_-  + \cq_-. 
                           \label{CanEq6}         
\end{eqnarray} 
Instead of the fourth-order Bach equations, we have now at our disposal the
nice differential system, given by Eq.~(\ref{CanEq1}) to Eq.~(\ref{CanEq6}),  
which is more appropriate for applying singularity analysis methods 
(Painlev\'e test) in order to extract all the exact solutions. In contrast 
with these methods which probe the local analytical structure of the canonical 
system, we can also seek out exact solutions by performing a global involution
algorithm on specific extra constraints chosen in accordance with local 
results from the analytic structure.

\subsection{Global involution of extra constraints}

The involution method consists in applying the Dirac--Bergmann consistency
algorithm on our system, with the Poisson brackets defined \wrt{} the 
canonical variables $\beta_{\pm}$, $\Pi_{\pm}$, $\cq_{\pm}$, $\cp_{\pm}$, and 
after suitable conditions have been imposed. Strictly speaking, the steps of 
the global involution algorithm are the following:
\begin{itemize}
   \item Impose an appropriate extra constraint on the canonical variables;
   \item Require that constraint to be preserved when time evolution is
         considered. This gives rise to secondary constraints and possibly to
         the determination of the Lagrange multiplier associated with the
         extra constraint; 
   \item Repeat the second step (involution) until no new information comes 
         out.
\end{itemize}
Once the involution algorithm has been performed we can classify all the
constraints into first class and second class and proceed further to the
analysis of the particular system. \\

Among the set of constraints we have considered in our analysis two are more
significant. The first expresses that the ratio of the variables $\cp_{\pm}$ 
is constant. Any solution to the canonical equations that satisfies that 
specific constraint is conformally equivalent to an Einstein space. 
Consistency of the extra constraint yields only one secondary constraint and
both are second class. We eliminate the associated spurious \dofs{} and reduce 
the canonical equations (\ref{CanEq3}--\ref{CanEq6}) to one binomial equation 
of Briot and Bouquet. The representations of the solution on the real axis are 
complicated expressions involving the Weierstrass elliptic function.%
~\cite{jdlqcs} As a particular case of that analysis we obtain the general 
axisymmetric solution. The second interesting constraint requires that the
momenta $\Pi_{\pm}$ be zero. In accordance with local results from the 
analytic structure, the general solution to the complete system with 
$\Pi_{\pm}=0$ is easy to produce under analytic form, upon integrating linear 
ODE's of the Lam\'{e} type. It provides a specific example of a solution to 
the Bach equations that is not conformally related to an Einstein space.%
~\cite{jdlqcs}
  
\section{Analytic structure of Bianchi type--I cosmology}
\label{sec:Painleve}

The existence of the above solutions, whether particular solutions of the
general differential system or general solutions of specialized systems, tells 
nothing about the integrability or non-integrability of the complete system 
and gives no information whatsoever about the mere accessibility of an exact 
and closed-form analytic expression of its general solution. This is due to 
the fact that the global involution algorithm of the extra constraints, as 
operated above, is not related with integrability and may even prove to be 
non-exhaustive. We have tackled the integrability issue through an invariant 
investigation method of intrinsic properties of the general solution.%
\footnote{For a circumstantial analysis, we refer the reader to our 
forthcoming paper.~\cite{jdlqcs}} 
In particular, we have proved analytically that the system under consideration 
is not integrable: its general solution exhibits, in complex time, an infinite 
number of logarithmic transcendental essential movable singularities --- \ie{}
an analytic structure not compatible with integrability in the practical 
sense; the quest for generic, exact and closed-form analytic expressions of 
the solution is hopeless. We stress that this result holds under spacetime 
transformations within the equivalence class of the Painlev\'e property.%
~\cite{conte} 

\section{Conformal relationship with Einstein spaces}
\label{sec:einspace}

The vanishing of the Bach tensor is a necessary condition for a Riemann space
to be conformal to an Einstein space. It means that any solution of vacuum
\gr{} or any space conformal to an Einstein space are also a solutions of 
conformal gravity. The converse however is not true: there exist spaces not 
conformally related to Einstein spaces that satisfy the Bach equations.%
~\cite{kozam} A necessary and sufficient condition for a space to be conformal 
to an Einstein space is the existence of a function $\sigma({\bf x})$, \ie{} 
conformal factor, that satisfies the differential equations
\begin{equation} \label{LEq1}
   L_{ab} = \nabla_a \sigma \nabla_b \sigma -
            \nabla_a \nabla_b \sigma -
            \frac12 g_{ab} g^{cd} \nabla_c \sigma \nabla_d \sigma -
            \frac{\tilde{R}}{24} \rme^{2 \sigma} g_{ab} = 0,
\end{equation}
where the tensor $L_{ab}$ is defined by $L_{ab}:=(R g_{ab} - 6 R_{ab})/12$.
For the Bianchi type--I model it is not difficult to show that the above
conditions in Eq.~(\ref{LEq1}) imply precisely that the ratio of variables
$\cp_{\pm}$ is constant (our first extra constraint in the algorithm). In that
case the conformal factor can be uniquely determined as a function involving
the Weierstrass elliptic function. Moreover it is possible to obtain the 
explicit form of the constant scalar curvature, $\tilde{R}$, of the conformal 
Einstein space. On the other hand, our solution obtained by imposing 
$\Pi_{\pm} = 0$ is not conformal to an Einstein space, for the ratio of 
variables $\cp_{\pm}$ is not constant in that case. We have thus confirmed
explicitly Schmidt's conjecture of the existence of what he calls `non-%
trivial' solutions to the Bach equations.~\cite{schmi}

\section*{Acknowledgments}

We are especially grateful to J. Demaret and C. Scheen for our collaboration
and stimulating discussions. We thank Dr Schimming for hints to the literature 
with regards to the Bach theory. We acknowledge financial support from the 
``Patrimoine de l'Universit\'e de Li\`{e}ge''. This work was supported in part 
by Contract No. A.R.C. 94/99-178 and by a F.R.I.A. grant.

\end{document}